\documentclass[preprint,showpacs,amsmath,amssymb]{revtex4-1}
\usepackage{mathrsfs}

\usepackage{graphicx,color,multirow}
\usepackage{dcolumn}
\usepackage{bm}
\newcommand{\nn}{\nonumber}

\newcommand{\MG}{ {\sc Madgraph5\_aMC@NLO} }
\newcommand {\beq} {\begin{equation}}
\newcommand {\eeq} {\end{equation}}
\newcommand {\bea} {\begin{eqnarray}}
\newcommand {\eea} {\end{eqnarray}}

\newcommand{\ppkkb}{$pp \to X_D \overline{X}_D + b + X$}

\begin{document}

%-------------------------------------------------------------------------------------------------------

\title{Higher-Order QCD prediction for dark matter pair associated with a b-jet production at the LHC}

\author{Li Gang$^a$}\email{lig2008@mail.ustc.edu.cn}
\author{Yu Si-He$^a$}%\email{yusihe234@163.com}
\author{Song Mao$^a$}%\email{songmao@mail.ustc.edu.cn}
\author{Zhang Yu$^b$}%\email{dayu@mail.ustc.edu.cn}
\author{Zhou Ya-Jin$^c$}%\email{zhouyj@sdu.edu.cn}
\author{Guo Jian-You$^a$}%\email{jianyou@ahu.edu.cn}

\affiliation{$^a$ School of Physics and Material Science, Anhui University, Hefei, Anhui 230039, P.R.China}
\affiliation{$^b$ School of Physics, Nanjing University, Nanjing, Jiangsu 210093, P.R.China}
\affiliation{$^c$ School of Physics, Shandong University, Jinan Shandong 250100, P.R. China }
\date{\today}

%-------------------------------------------------------------------------------------------------------
\begin{abstract}
Dark matter associated visible particle production at high energy colliders provides a unique
way to determine the microscopic properties of the dark matter. We investigate a pair of fermionic dark
matter particles associated with a b-jet production at the LHC, through a mediator which couples to standard model
or dark matter particles via either a vector or axial-vector coupling. The calculation is performed by implementing these simplified models in the
FeynRules/MadGraph5 aMC@NLO framework. In our calculation, next-to-leading order
QCD corrections and parton-shower effects are considered. We find that this process has a sizeable cross section and the QCD correction
can reach more than 2 times than LO results. We also investigate the discovery potential in several benchmark scenarios at the 13TeV LHC.
\end{abstract}

\maketitle

\section{Introduction}
\par
The astrophysical and cosmographic observational evidences have
confirmed the existence of dark matter (DM) and provided the DM density
in our universe \cite{Bertone04}. However, these
observations don't tell us the information about the mass of DM particle or
whether it interacts with the Standard Model (SM) particles.
Determining the nature of DM particle quantitatively is one of the
most important tasks both in cosmology and particle physics. Among all
the DM candidates, weakly interacting massive particles (WIMPs) is
the most compelling one. This is due to that it offers the
possibility to understand the relic abundance for the DM as a
natural consequence of the thermal history of the universe
\cite{Feng2008a}. Some extensions of the SM, such as
Supersymmetry \cite{Martin:1997ns,Drees:2004jm}, Universal Extra
Dimensions \cite{Appelquist:2000nn} or Little Higgs Models
\cite{ArkaniHamed:2001nc,Wu:2016rwz}, naturally lead to good candidates for
WIMPs and the cosmological requirements for the WIMP abundance in
the universe. However, all of these theories still lack experimental
support, and it is difficult to judge which theory is proper for the
DM particle. Additionally, the first observation of the DM may come
from direct- or indirect-detection experiments, which is difficult
to provide information about the general properties of the DM
particle. Thus, model-independent studies of DM phenomenology are particularly important.

\par
There are many experiments currently running or planned aiming to
test the hypothesis by searching for WIMPs. These experiments can
be divided into two classes: direct detection experiments, which
search for the scattering of dark matter particles off atomic nuclei
within a detector, such as CDMS, XENON, LUX, PandaX; and indirect
detection experiments, which look for the products of WIMP annihilations, such
as Fermi Gamma-ray Space Telescope, PAMELA. An alternative approach
to the detection of WIMPs in nature is to produce them at high
energy colliders. Experiments with the Large Hadron Collider (LHC)
may be able to detect WIMPs produced in collisions of the LHC proton
beams. Because a WIMP has negligible interactions with matter, it will
may not be detected as missing energy and momentum which
escapes the LHC detectors, the useful method is searching the DM
particle production associated a visible particle, such as jet,
gauge boson, heavy quark.

\par
Recently, some observational results favour a light DM with a mass
around $10~ {\rm GeV}$ in various experiments. The DAMA experiment
has reported a signal of annual modulation at a highly significant
level \cite{Bernabei:2010mq}, which is consistent with a discovery
interpretation from a low mass dark matter in direct measurements by
CoGeNT~\cite{cogent}, CRESST~\cite{cresst} and
CDMS~\cite{Agnese:2013rvf} experiments. While this region of
parameter space is excluded by the other experiments like XENON100
\cite{Aprile2012}, LUX \cite{Akerib2014} and SuperCDMS
\cite{Agnese2014}. In order to clarify this puzzle, there have been
more researches in light DM models (where the DM mass is
order of a few GeV)
\cite{Kim:2009ke,Fitzpatrick:2010em,Kopp:2009qt,Kuflik:2010ah,Chang:2010yk,Essig:2010ye,An:2010kc,Andreas:2010dz,
Barger:2010yn,Hooper:2010uy}.

\par
In the case of a WIMP, stability on the order of the lifetime of the
universe implies that pair production must highly dominate over
single production, and precludes the WIMP from decaying within the
detector volume. Searches for dark matter in missing momentum
channels can be classified based on the visible particles against
which the invisible particles recoil. Existing experimental studies
have considered the cases in which the visible radiation is a jet of
hadrons (initiated by a quark or
gluon)~\cite{cdfjmet,ATLAS:2012ky,Chatrchyan:2012me}, a
photon~\cite{Aad:2012fw,Chatrchyan:2012tea}, or a $W/Z$ boson
decaying into leptons or hadronic jets
~\cite{TheATLAScollaboration:2013fia,CMS:2013iea,Aad:2013oja}.
The bottom quark can be identified by reconstructing secondary vertices, and the high-$p_T$ bottom quark can be tagged with reasonably high efficiency at the LHC, meanwhile the observation of a bottom quark with high-$p_T$ can reduce the backgrounds of the dark matter production. Thus, it is very interesting to study the dark matter pairs associated with a (anti)bottom quark production at the LHC. Because the LHC is a proton-proton collider, the QCD correction
should be considered for any process if someone wants to make a
reliable prediction. More recently, the production of DM pairs plus
a jet, photon and $W/Z$ have been calculated to QCD next-to-leading order
(NLO) \cite{Backovic:2015soa,Haisch:2012kf, Fox:2012ru,Huang:2012hs,Wang:2011sx,Haisch:2013ata,Neubert:2015fka,Haisch:2016usn,Mattelaer:2015haa,Chen:2016ylf}.
In Ref.\cite{Lin:2013sca}, the DM pairs associated with a bottom quark production at LO for scalar coupling have been studied. In this work, we investigate a pair of fermionic dark matter particles associated with a b-jet production up to QCD NLO in the simplified model at the LHC.

\par
The paper is arranged as follows: in Section II we briefly describe
the related simplified model and present the
calculation strategy. In Section III, we present some numerical
results and discussion. Finally, a short summary is given in Section IV.

\vskip 5mm
\section{Simplified model and calculation strategy }
\par
We assume that the dark matter candidate is a new particle
which is singlet under the SM local symmetries, and all SM particles
are singlets under the dark-sector symmetries. The interactions
between the SM and DM sectors are presumably effected by the
exchange of some heavy mediators, which could be a vector, axial vector or scalar particles.
Given the assumption that the WIMPs are SM singlets, the
factor in each operator consisting of SM fields must also be
invariant under SM gauge transformations. The interactions between
the DM and SM quarks are described by the simplified model in
Refs\cite{Mattelaer:2015haa,Backovic:2015soa,Neubert:2015fka}.

In the framework of the simplified model, the interaction  of
a spin-1 vector or axial-vector mediator ($Y_1$) with a Dirac
fermion DM ($X_D$) is given by
 \begin{align}\label{eq:vector_mediator}
 {\cal L}_{X_D}^{Y_1} = \bar X_D \gamma_{\mu}
 (g^{V}_{X_D}+g^{A}_{X_D}\gamma_5)X_D\,Y_1^{\mu} \,,
 \end{align}
 and with quarks by
 \begin{align}\label{eq:vector_mediator2}
  {\cal L}_{\rm SM}^{Y_1} &= \sum_{i,j} \Big[\bar d_i\gamma_{\mu}
  (g^{V}_{d_{ij}}+g^{A}_{d_{ij}}\gamma_5)d_j \nn\\
  &\hspace*{10mm} +\bar u_i\gamma_{\mu}
  (g^{V}_{u_{ij}}+g^{A}_{u_{ij}}\gamma_5)u_j\Big] Y_1^{\mu} \,,
  \end{align}
 where $u$ and $d$ represent up- and down-type quarks, respectively.
 $g^{V/A}$ are the vector/axial-vector
 couplings of DM and quarks, and  $i,j$($i,j$=1,2,3) are flavour indices.
 This notation are adopted to the actual implementation
 in {\sc FeynRules}~~\cite{Alloul:2013bka}. The model files can be downloaded at the {\sc FeynRules} website~\cite{FR-DMsimp:Online}.

The pure vector and pure axial-vector mediators are given by
setting the parameters in the Lagrangians~\eqref{eq:vector_mediator} and
\eqref{eq:vector_mediator2} to
\begin{align}
&g^V_{X_D} \equiv g_X \quad {\rm and}\quad g^A_{X_D} = 0
\label{paramX_v} \\
&g^{V}_{u_{ii}} =  g^{V}_{d_{ii}} \equiv g_{\rm SM} \quad {\rm and}\quad
g^{A}_{u_{ii}} =  g^{A}_{d_{ii}} = 0
\label{paramSM_v}
\end{align}
 and
\begin{align}
&g^V_{X_D} = 0 \quad {\rm and}\quad g^A_{X_D} \equiv g_X
\label{paramX_a}\\
&g^{V}_{u_{ii}} = g^{V}_{d_{ii}} = 0 \quad {\rm and}\quad
g^{A}_{u_{ii}} = g^{A}_{d_{ii}} \equiv g_{\rm SM}\,,
\label{paramSM_a}
\end{align}
respectively, where we assume quark couplings to the mediator are the same for all the
flavours and set all flavour off-diagonal couplings to zero.
With this simplification of a single universal coupling for the SM-$Y_1$
interactions, the model has only four independent parameters, $i.e.$ two couplings
and two masses:
\begin{equation}\label{param}
\{g_{\rm SM},\,g_X,\,m_X,\,m_Y\} \,.
\end{equation}
We note that the mediator width is calculated from the above
parameters. In the following sections, we take $g_{\rm SM} = 1$ and $\,g_X = 0.25$  as our benchmark for the vector
and axial-vector mediator scenario~\cite{Backovic:2015soa,Neubert:2015fka}.

\begin{figure}[!htb]
\begin{center}
{\includegraphics[width=10cm]{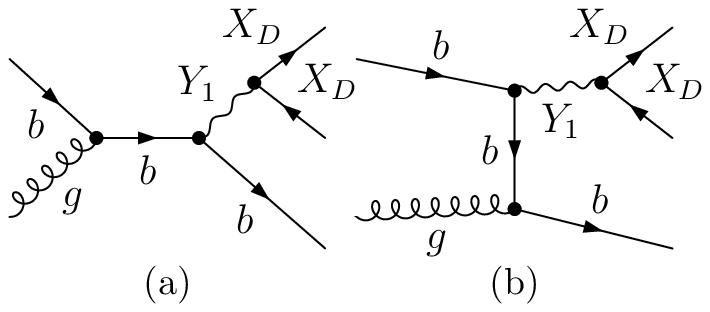}}
\end{center}
\vspace*{-0.5cm}\caption{\label{fig1} The LO Feynman diagrams for the process \ppkkb ~at the LHC.}
\end{figure}

As we know, the cross section for the partonic process $g(p_1) + b(p_2) \to X_D(k_3) \overline{X}_D(k_4) + b(k_5)$
should be the same as that for its charge conjugate subprocess $g(p_1) + \bar{b}(p_2) \to X_D(k_3) \overline{X}_D(k_4) + \bar{b}(k_5)$,
and the luminosity of the bottom quark in a proton is same as the anti-bottom quark.
Therefore, the production rates of the $X_D \overline{X}_D b$ and the $X_D \overline{X}_D\bar{b}$ are identical at the LHC.
In the following sections, we denote that the parent process $p + p \to X_D \overline{X}_D + b + X$ includes the two partonic process above unless otherwise indicated. There are two Feynman diagrams for this process at leading order(LO), which are shown in Fig.\ref{fig1}.
In our calculation, the mediator width is automatically computed by using
the MadWidth module for each parameter point.
The above benchmark coupling strength leads to
$\Gamma_Y/m_Y\sim0.05$ for $m_Y > 2m_X$ and $\Gamma_Y/m_Y\sim0.025$ for
$m_Y < 2m_X$ both for the vector and axial-vector mediators.
Our calculation and simulation are based on the framework of {\sc MadGraph5$\_$aMC@NLO} \cite{Alwall:2014hca}.  The one-loop QCD corrections are computed using the program  {\sc MadLoop} \cite{Hirschi:2011pa}£¬which is based on the OPP method \cite{Ossola:2006us,Ossola:2007ax}.  The ultraviolet divergences and rational $R_2$ terms are calculated automatically using the {\sc NLOCT} package \cite{Degrande:2014vpa} and the infrared subtraction terms for real emissions are generated by {\sc MadFKS} \cite{Frederix:2009yq}. Event generation is obtained by matching
short-distance events to the parton shower using the {\sc MC@NLO} framework \cite{Frixione:2002ik}, which is implemented for Pythia6 \cite{pythia6}.
We define all jets, including b-jet, using the anti-$k_T$ algorithm as implemented in FastJet with the
jet cone radius $R = 0.4$. Additionally, we require $p_T (j) > 15~GeV$ and $\eta(j) < 4.5$ for all jets in the event.
The b-jets, originating from b-quarks, are defined as the long lifetime and the large mass of b hadrons.

\vskip 5mm
\section{Numerical results and discussion}
\subsection{Total cross section}

\par
In this section we provide LO and NLO QCD predictions for the total cross sections for the process \ppkkb ~at the 13 TeV LHC.
We take NN23LO1 and NN23NLO PDF sets \cite{Ball:2013hta} for LO and NLO calculations, and the
corresponding fitted values $\alpha_s(M_Z) = 0.130$ and
$\alpha_s(M_Z) = 0.118$ are used for the LO and NLO calculations,
respectively. The central value $\mu_0$ for the renormalisation ($\mu_R$) and factorisation ($\mu_F$) scales is set to $H_T/2$, where $H_T=\sum_i\sqrt{p_{T,i}^2+m_i^2}$, which is the sum of the transverse momenta of all final-state particles. The scale uncertainty is estimated by varying the scales $\mu_R$ and $\mu_F$, from $\mu_0/2$ to $2 \mu_0$, independently.

\par
In table \ref{tab1}, we list the total cross sections of dark matter associated with various visible particles production at 13 TeV LHC, which is obtained with \MG. For all the jets, including b-jet, we set the distance in the $(\eta, \phi)$ plane $R = 0.4$, $p_T (j) > 15~GeV$ and $\eta(j) < 4.5$ . For the process $p p \to X_D \overline{X}_D + \gamma + X$, we set the $p_T (\gamma) > 20~GeV$. For the process $p p \to X_D \overline{X}_D W^{\pm} + X$ and $p p \to X_D \overline{X}_D Z + X$, we don't consider the $W^{\pm}$ and $Z$ subsequent decay and add any select cuts. We find that the cross section of $p p \to X_D \overline{X}_D + b + X$ is comparable with the process $p p \to X_D \overline{X}_D + \gamma + X$, much lager than the process $p p \to X_D \overline{X}_D W^{\pm} + X$ and $p p \to X_D \overline{X}_D Z + X$. If considering the $W$ and $Z$ boson subsequent leptonic or hadronic decay, the cross section for the process $p p \to X_D \overline{X}_D W^{\pm} + X$ and $p p \to X_D \overline{X}_D Z + X$ will be even smaller.
\begin{table*}
\begin{footnotesize}
\center
\begin{tabular}{|l|l|l|l|l|l|l|}
\hline
  & $X_D\overline{X}_Dj$ & $X_D\overline{X}_D\gamma$ & $X_D\overline{X}_D
  b$ & $X_D\overline{X}_DZ$ & $X_D\overline{X}_DW^{\pm}$\\
  \hline
  $\sigma(pb)$ & 8.245$\pm$ 0.053& 0.1202$\pm$ 9.0e-4 & 0.1019$\pm$ 1.5e-3 & 0.04616$\pm$1.2e-4 & 4.796e-3$\pm$1.3e-5 \\
  \hline
\end{tabular}
\caption{The NLO cross sections for DM pair production in association with a visible particle for the vector mediator at the 13 TeV
LHC with the coupling parameters $g_{X}=1$ and $g_{\rm SM}=0.25$.}
\label{tab1}
\end{footnotesize}
\end{table*}

\par
\begin{figure}
\begin{center}
\includegraphics[bb = 25 225 490 565,scale = 0.45]{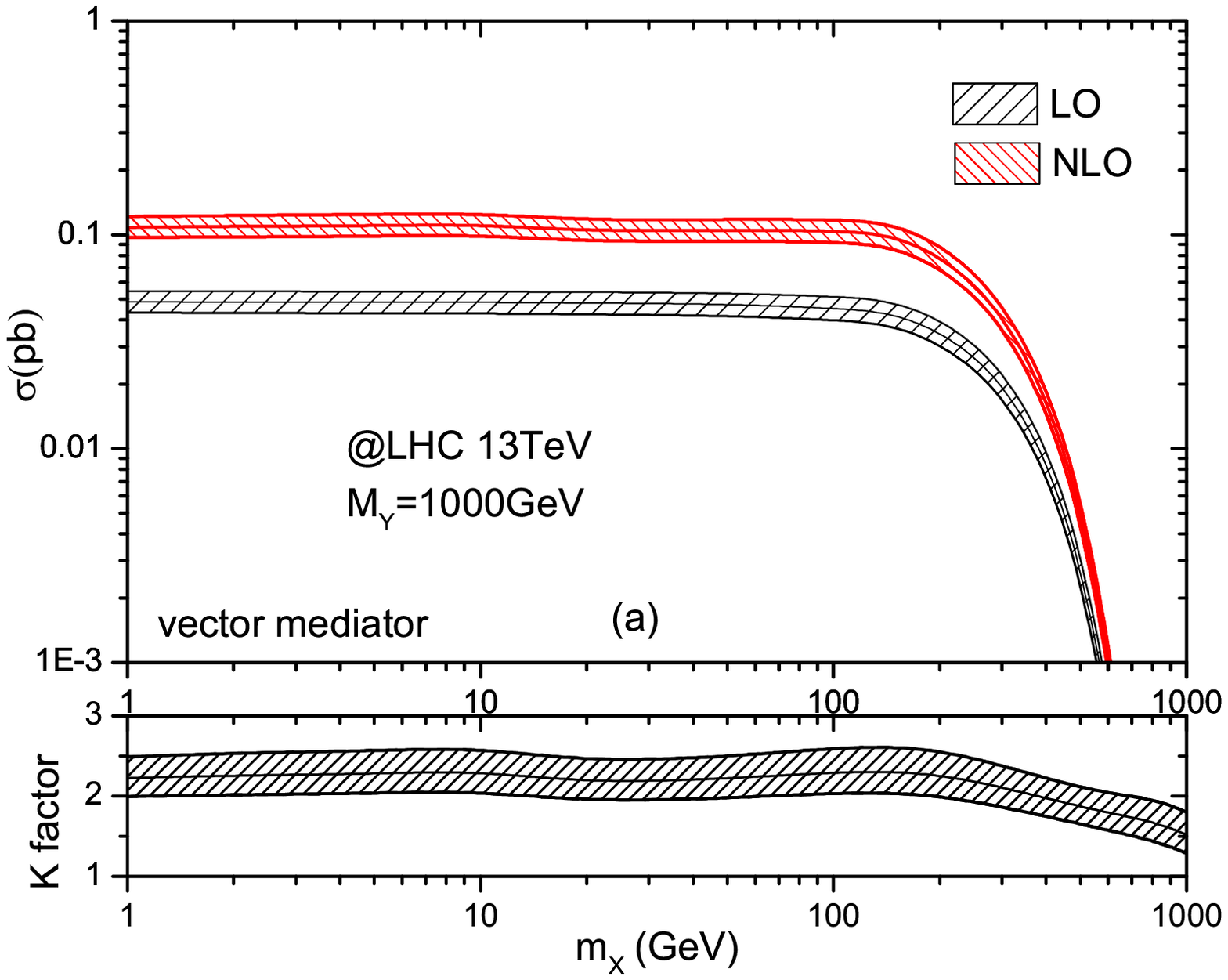}
\includegraphics[bb = 25 225 490 565,scale = 0.45]{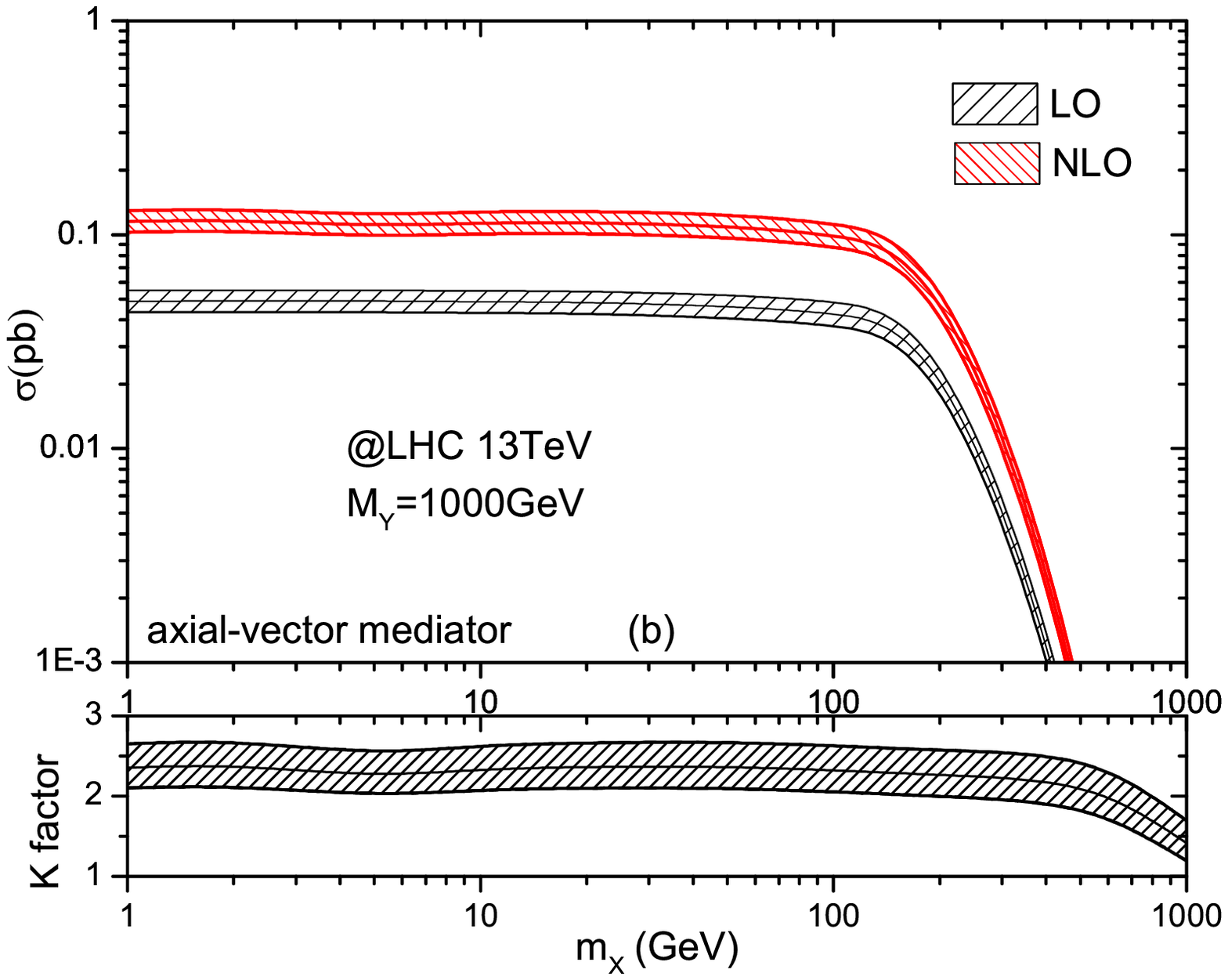}
\vspace*{0cm}
\caption{\label{fig2} The LO, NLO QCD corrected integrated cross sections and the corresponding $K$-factors as the functions of the DM mass for
the process \ppkkb~ at the $13~TeV$ LHC with $M_{Y}=1000~GeV$ and $\mu=\mu_0$,
the shaded band represents the deviation from this scale when the scales vary from  $\mu_0 /2 $ to $2 \mu_0$.}
\end{center}
\end{figure}

In Fig.\ref{fig2}, we present the DM mass dependence of the cross
sections and the corresponding K-factors for the \ppkkb~ process
induced by the vector, axial-vector mediator by taking
$M_{Y}=1000~ {\rm GeV}$ at 13TeV LHC, separately.
The dependence on the renormalization and factorization scales, with the simplification
$\mu=\mu_r=\mu_f$, is illustrated by the shaded band linking the
predictions obtained at $\mu=2 \mu_0$ and $\mu=1/2 \mu_0$, while
the central scale choice $\mu=\mu_0$ is illustrated by the curve
inside the shaded band. As shown in the figures, the
cross sections are insensitive to the DM mass $m_{X}$ in the range
of $m_{X} < 100~ {\rm GeV}$, and start to decreases rapidly with the
increment of $m_X$ when $m_{X} > 100~ {\rm GeV}$. This is due
to the final state phase space reduced rapidly when the dark matter mass $m_X$ increasing. We find that the contributions
from the vector-mediator and axial-vector-mediator can not be distinguished until $m_{X} > 100~ {\rm GeV}$.
The scale uncertainties are not significantly reduced when comparing the LO and NLO QCD predictions.
We define the K-factor as $ K(\mu) = \sigma^{(NLO)}(\mu)/\sigma^{(LO)}(\mu_0)$ and find that the K-factors are more than 2 at lower mediator mass for both the vector and axial-vector mediators.

\begin{figure}
\begin{center}
\includegraphics[bb = 25 225 490 565,scale = 0.45]{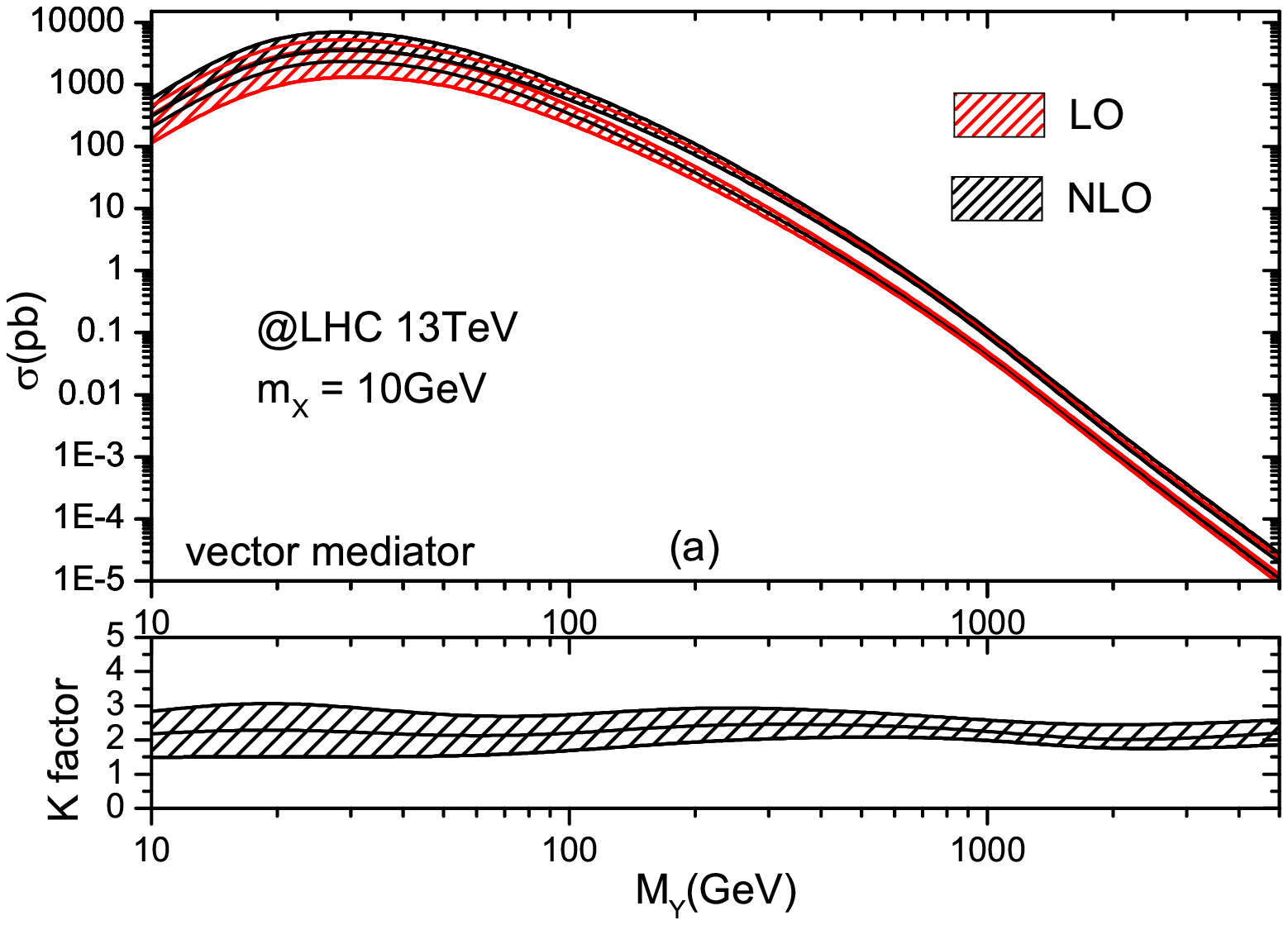}
\includegraphics[bb = 25 225 490 565,scale = 0.45]{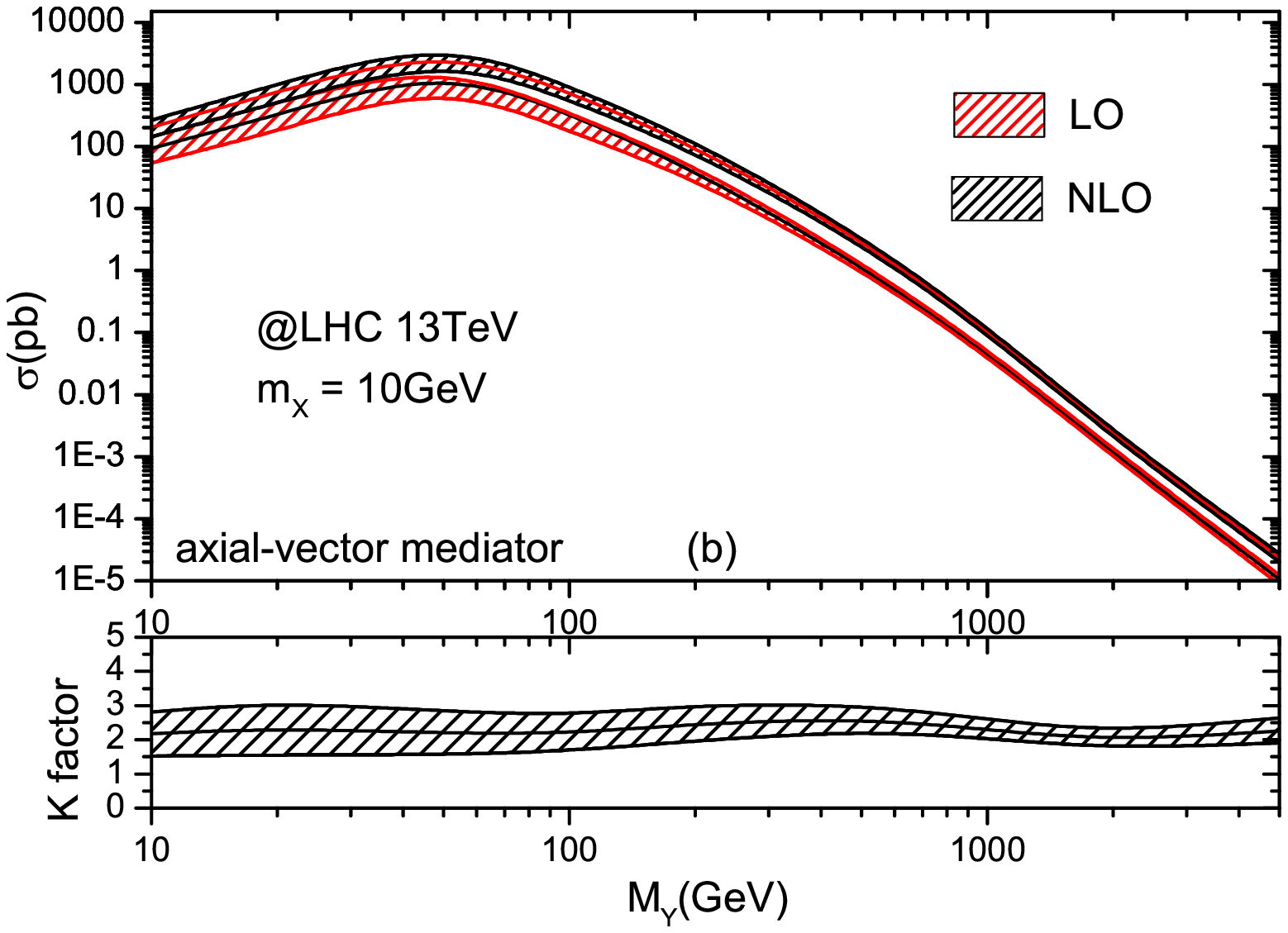}
\vspace*{0cm}
\caption{\label{fig3} The LO, NLO QCD corrected integrated cross sections and the corresponding $K$-factors as
the functions of the vector mediator mass for
the process \ppkkb~ at the $13TeV$ LHC with $m_{X}=10~GeV$ with the default scale $\mu=\mu_0$,
the shaded band represents the deviation from this scale when the scales vary from  $\mu_0 /2 $ to $2 \mu_0$.}
\end{center}
\end{figure}

\par
In Fig.\ref{fig3}, we show that the vector and axial-vector mediators mass dependence of the LO,
NLO QCD corrected integrated cross sections and the corresponding
K-factor for the process \ppkkb~ at the $13TeV$ LHC with $m_{X}=10~GeV$.
The dependence on the renormalization and factorization scales is illustrated by the shaded band linking the
predictions obtained at $\mu=2 \mu_0$ and $\mu=1/2 \mu_0$ with the simplification
$\mu=\mu_r=\mu_f$,  and
the central scale choice $\mu=\mu_0$ is illustrated by the curve
inside the shaded band. When the vector and axial-vector mediators mass varies from $10~GeV$ to
$1000~GeV$, the NLO QCD corrections modify the LO cross sections
obviously.  The similar behavior is demonstrated
in the monojet production at the LHC \cite{Fox:2012ru}. The vector
and axial-vector DM mediators show similar behaviour in terms of
K-factors and scale dependence. This is because that in the massless
limit the only terms which are sensitive to the axial nature of the
coupling are the four-quark amplitudes, which is a small part of
the total NLO cross section. The total cross sections and
K-factors of some benchmark points are shown in table \ref{tab2}.

\begin{table*}
\begin{footnotesize}
\center
\begin{tabular}{|l|l|l|l|l|l|l|}
\hline
 \multirow{2}{*}{$(m_Y,m_X) [GeV]$} & \multicolumn{3}{c|}{vector mediator} & \multicolumn{3}{c|}{axial-vector mediator} \\
\cline{2-7}
  & LO(pb) & NLO(pb) & K-factor & LO(pb) & NLO(pb) & K-factor \\
  \hline
  (10,1) & 1.778e4$\pm$ 58& 6.744e4$\pm$ 790 & 3.772 & 1.733e4$\pm$56 & 6.449e4$\pm$55  &3.721 \\
  \hline
  (10,10) & 204.8$\pm$ 0.7 & 443.5$\pm$ 5& 2.166 & 94.24$\pm$0.34 & 196.5$\pm$2.2 & 2.085 \\
  \hline
  (100,1) & 361.5$\pm$ 1.6 & 861.2$\pm$ 9.3 & 2.382 & 361$\pm$1.8 & 886.4$\pm$8.9 & 2.455 \\
  \hline
  (100,10) & 362.5$\pm$1.7 & 841.1$\pm$ 8.4 & 2.320 & 348.1$\pm$1.6 & 848.5$\pm$9.7 & 2.438 \\
   \hline
  (100,100) & 0.365 $\pm$1.3e-3 & 0.9901$\pm$8.8e-3 & 2.713 & 0.114$\pm$3.5e-4 & 0.2974$\pm$6.3e-3& 2.609 \\
  \hline
  (500,10) & 1.186$\pm$5.3e-3 & 2.695$\pm$3.1e-2 & 2.272 & 1.23$\pm$ 5e-3 & 3.572$\pm$3.7e-2 & 2.904 \\
  \hline
  (500,100) & 1.132$\pm$4.7e-3 & 3.245$\pm$3.9e-2 & 2.867 & 1.005$\pm$ 4e-3 & 2.746$\pm$0.031 & 2.732 \\
    \hline
  (1000,10) & 4.836e-2$\pm$ 2e-4 & 0.1019$\pm$1.5e-4 & 2.107 & 4.91e-2$\pm$2.2e-4 & 0.1173$\pm$1.4e-3 & 2.390 \\
  \hline
  (1000,100) & 4.588e-2$\pm$2e-4 & 0.1296$\pm$1.5e-3 & 2.825 & 4.35e-2$\pm$1.9e-4 & 0.1138$\pm$1.32e-3 & 2.616 \\
  \hline
  (1000,500) & 9.069e-3$\pm$3.2e-5 & 2.684e-2$\pm$6.2e-4 & 2.960 & 5.4e-4$\pm$2e-6 & 1.21e-3$\pm$1.2e{-5} & 2.241 \\
  \hline
  (1000,1000) & 2.53e-6$\pm$8.2e-9 & 4.91e-6$\pm$7.8e-8 & 1.941 & 4.77e{-7}$\pm$1.4e{-9} & 6.705e{-7}$\pm$6.7e{-9} & 1.406 \\
  \hline
\end{tabular}
\caption{LO and NLO QCD cross sections and corresponding $K$ factors for
 DM pair production in association with a b-jet for the vector and axial-vector mediators at the 13TeV LHC.
 Several benchmark points for the vector and axial-vector mediators and
 DM masses are presented with the coupling parameters $g_{X}=1$ and
 $g_{\rm SM}=0.25$ .}
\label{tab2}
\end{footnotesize}
\end{table*}

\subsection{Kinematic distributions}

\begin{figure}
\centering
\includegraphics[bb = 25 225 490 565,scale = 0.42]{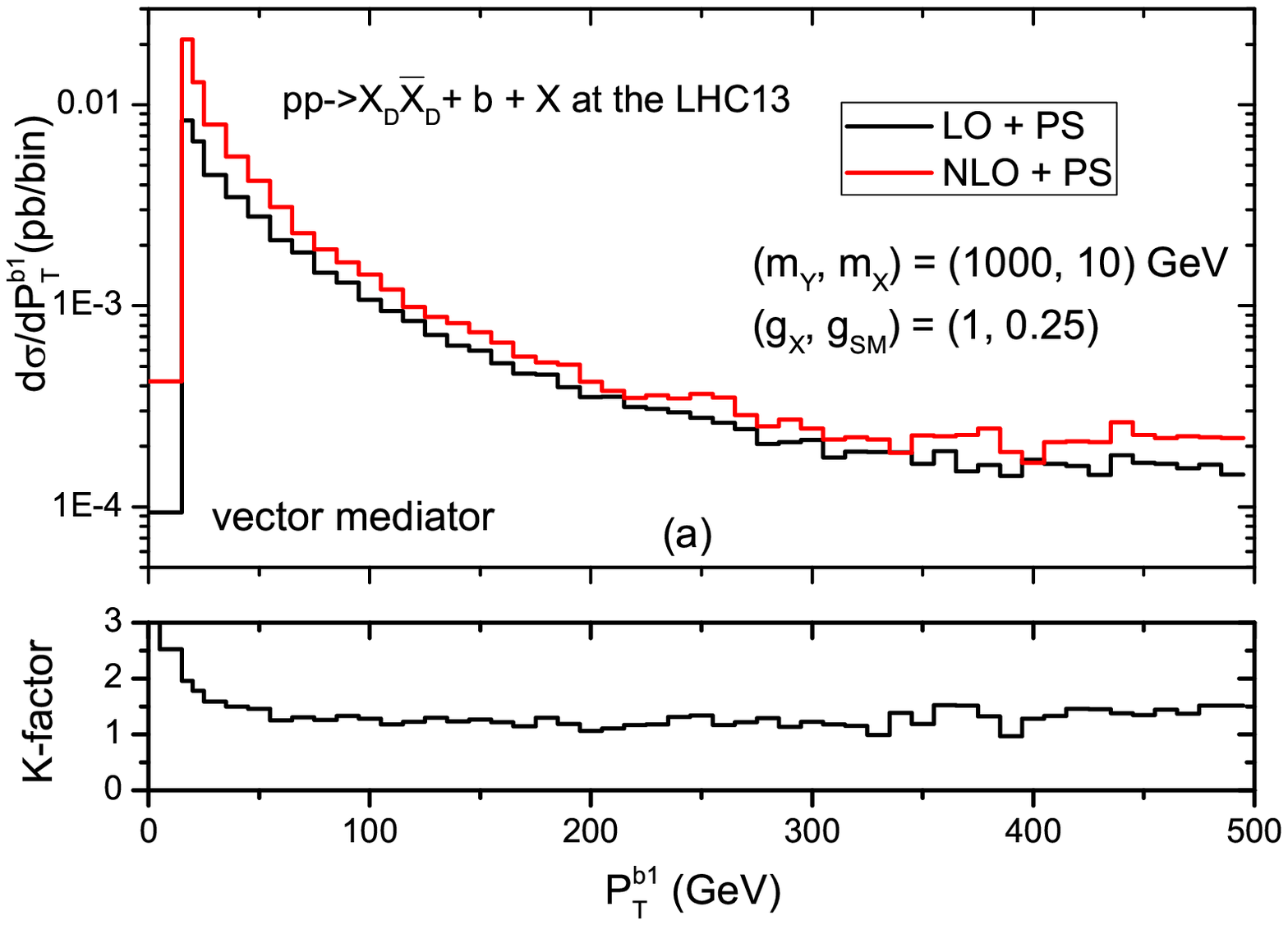}
\includegraphics[bb = 25 225 490 565,scale = 0.42]{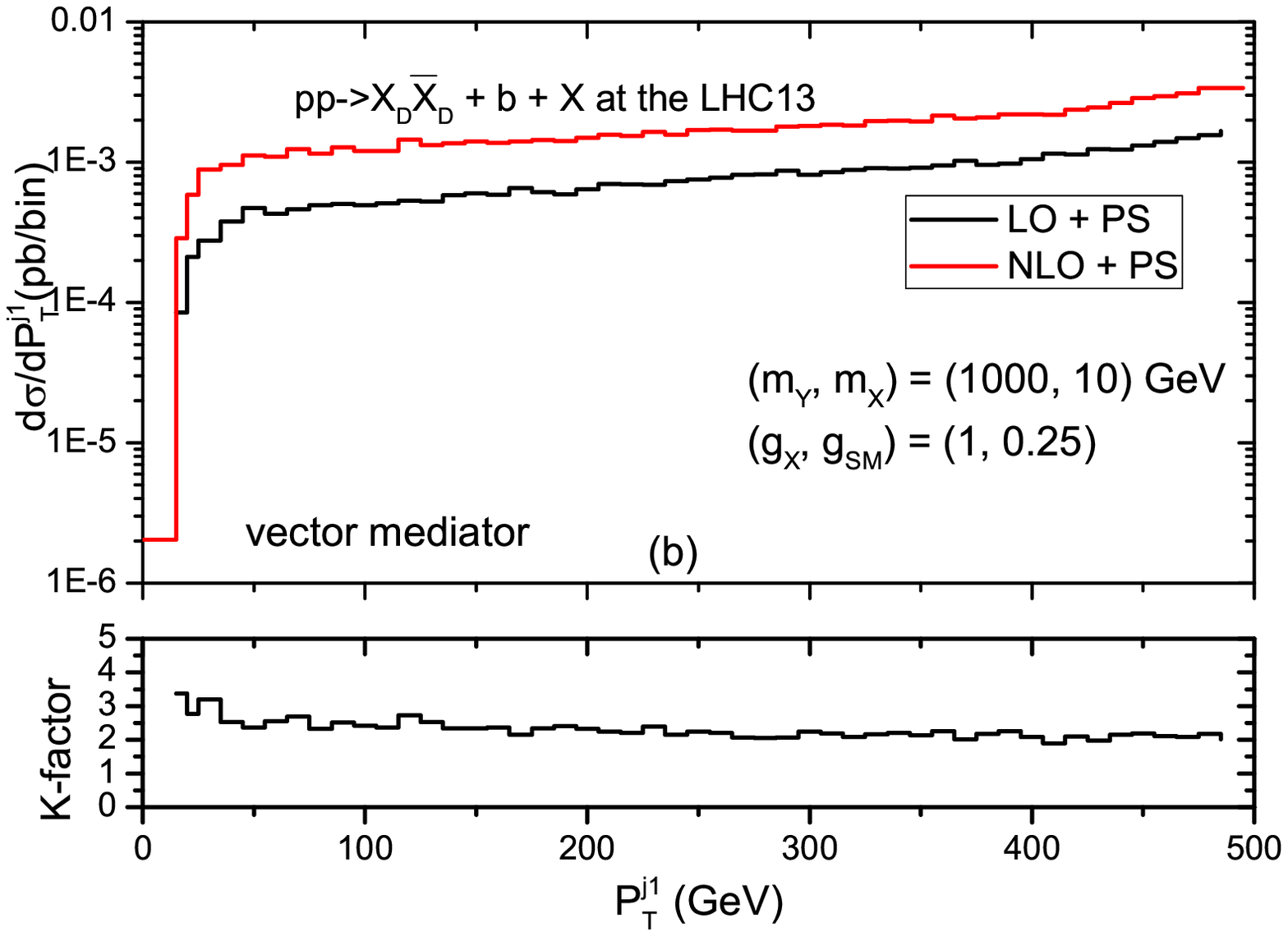}
\includegraphics[bb = 25 225 490 565,scale = 0.42]{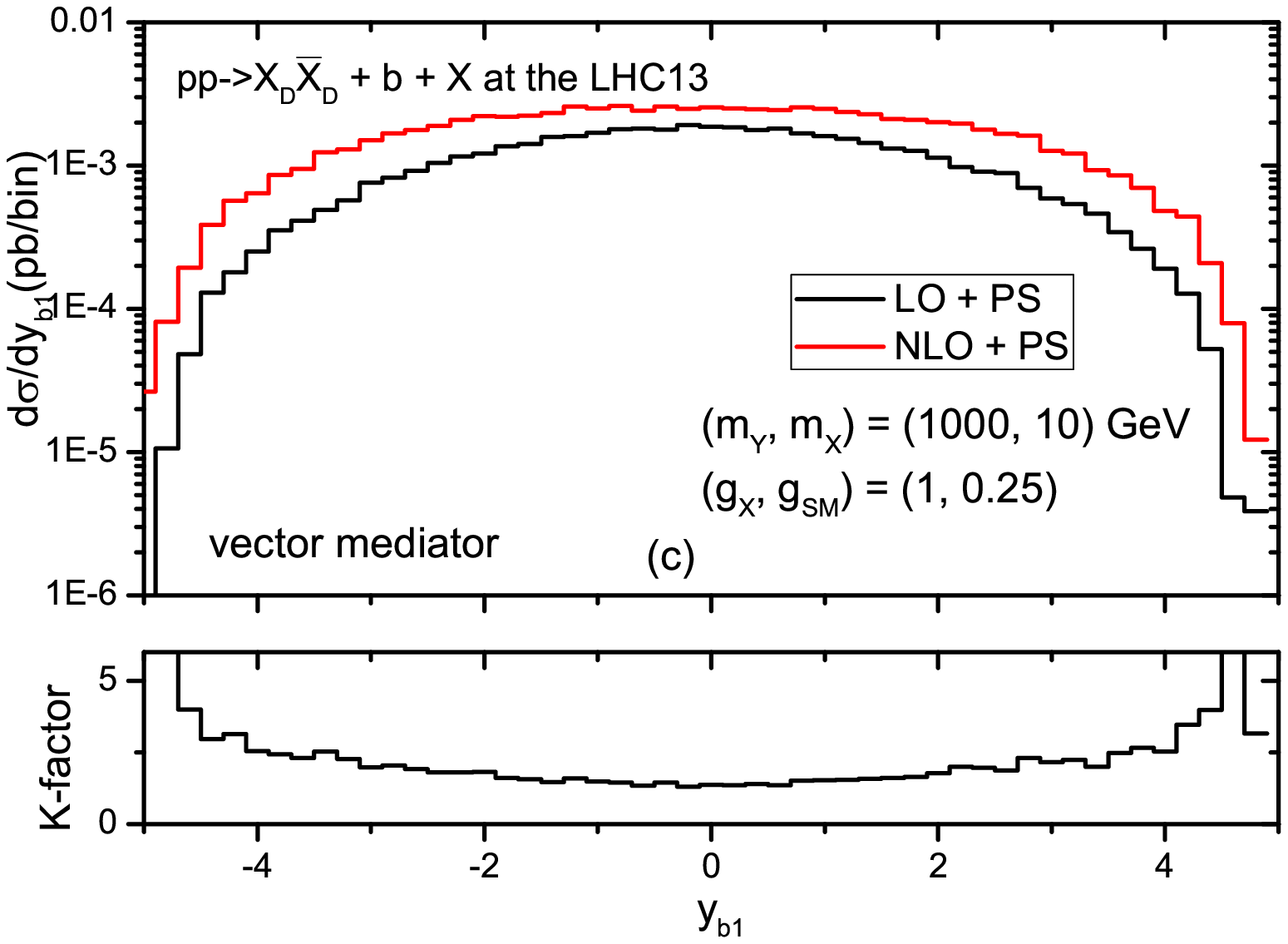}
\includegraphics[bb = 25 225 490 565,scale = 0.42]{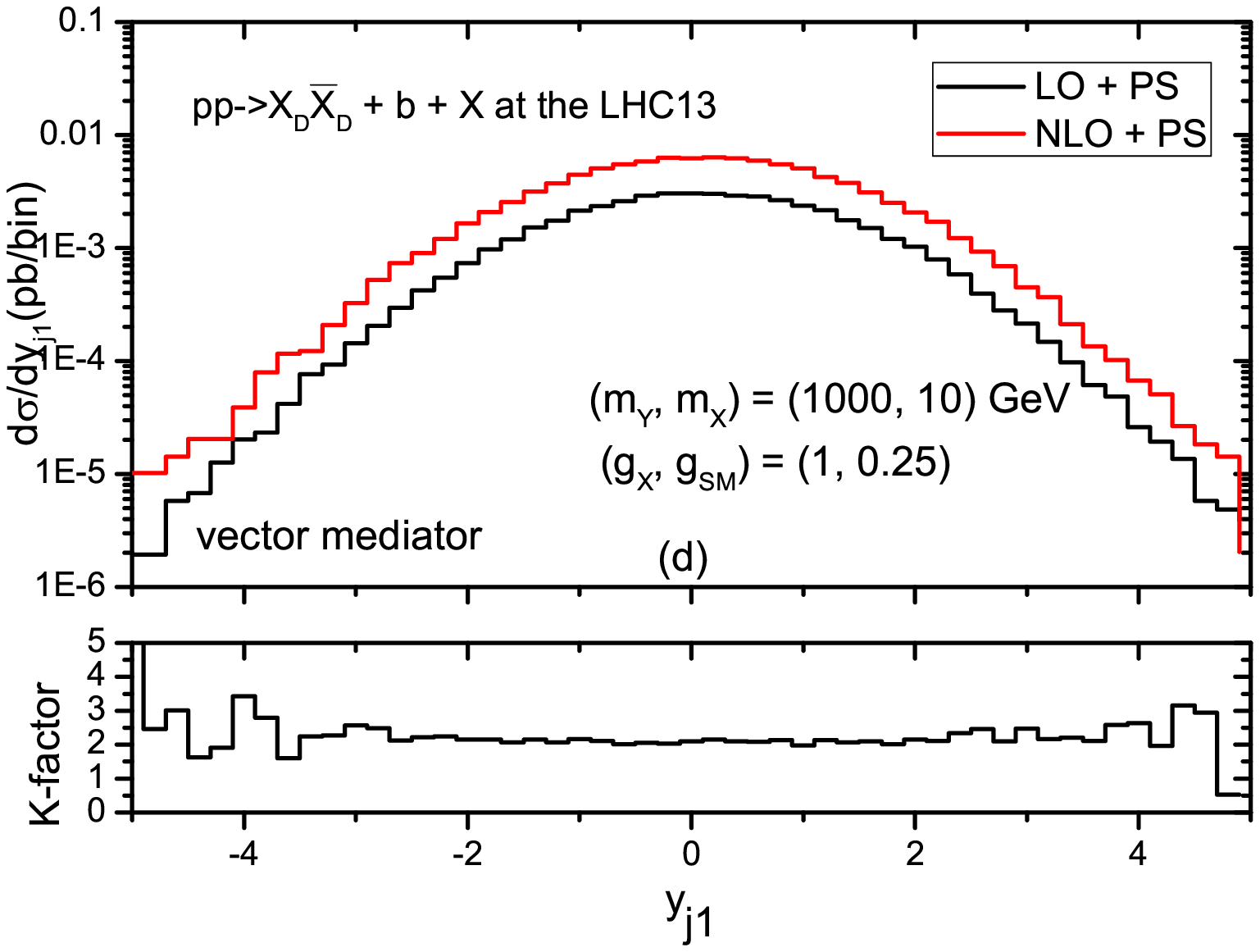}
\includegraphics[bb = 25 225 490 565,scale = 0.42]{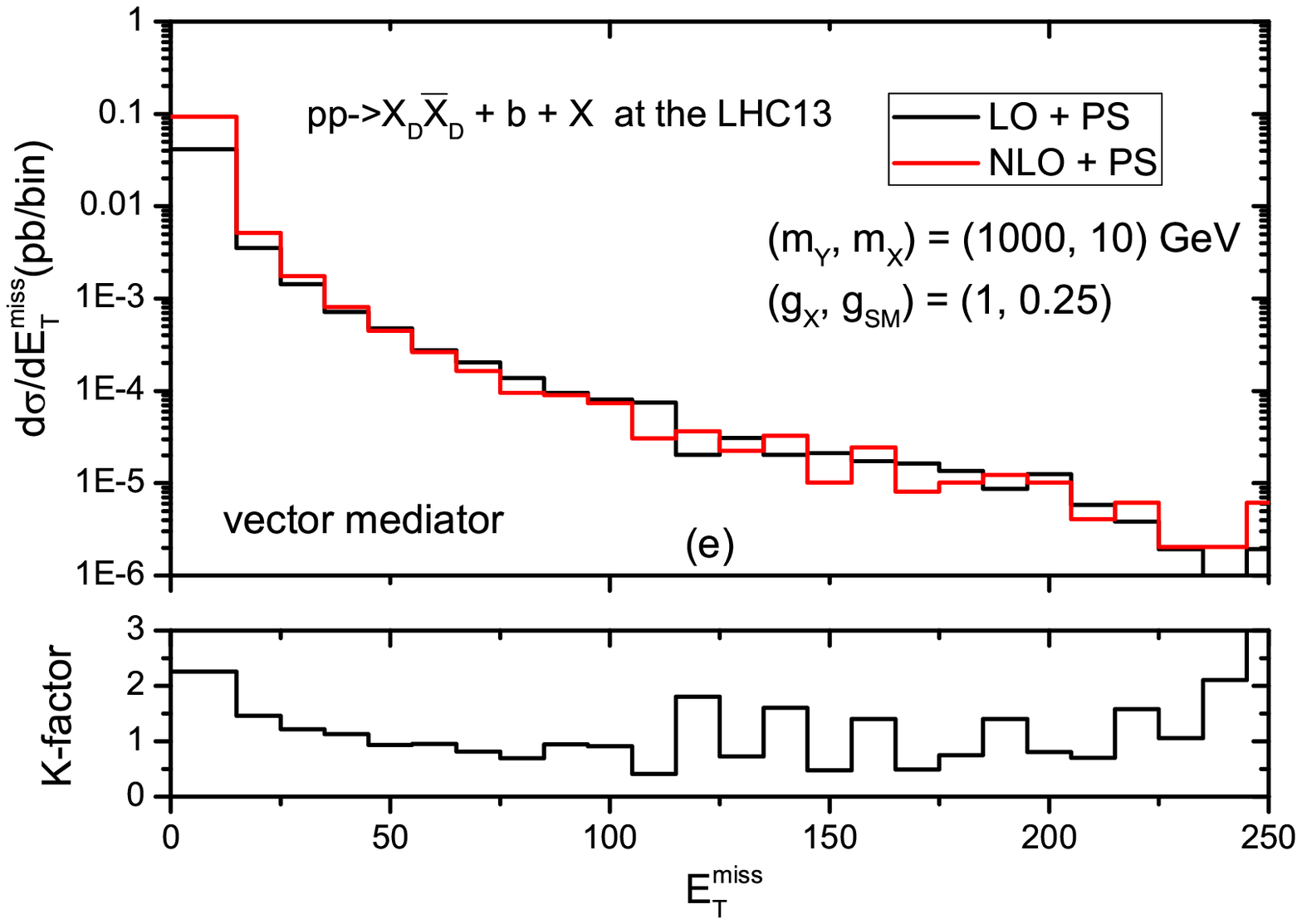}
\includegraphics[bb = 25 225 490 565,scale = 0.42]{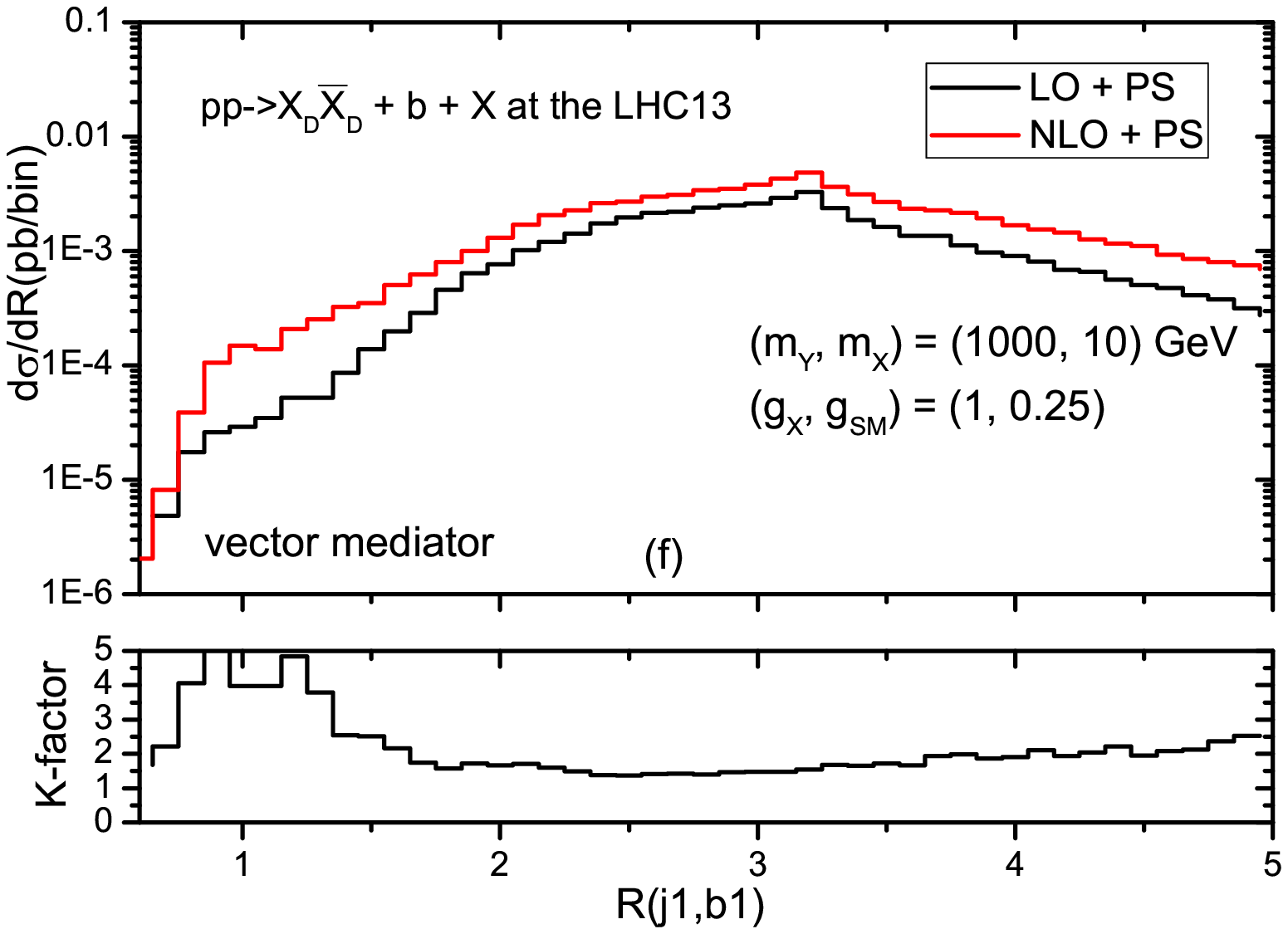}
\caption{\label{fig4} The LO, NLO QCD corrected distributions and corresponding $K$ factors of
the transverse momenta and rapidity of the hardest jet(j1), hardest b-jet(b1), the distributions of missing transverse momenta $E_T^{miss}$ and distance in the $(\eta, \phi)$ plane $R$ between the hardest jet(j1) and hardest b-jet(b1) for the \ppkkb~ processes at the $\sqrt{s}= 13TeV$ LHC. }
\end{figure}

\par
In order to obtain observable results in physics, we need to consider the parton-shower effects for the process \ppkkb.
We use Pythia6 to perform the parton-shower simulations. In Fig.\ref{fig4}, we provide the differential distributions and the corresponding $K$ factors of transverse momenta $p_T$ and rapidity $y$ for the hardest jet(j1) and hardest b-jet(b1), the distance in the $(\eta, \phi)$ plane $R$ between the hardest jet(j1) and hardest b-jet(b1), and the missing transverse momenta $E_T^{miss}$ at 13TeV LHC with $m_Y = 1000 ~GeV$ and $m_X = 10 ~GeV$ for vector mediator. For the $p_T$ differential distributions of the hardest b-jet(b1) in Fig.\ref{fig4}(a), the NLO QCD correction enhance the LO results at all region significantly, especially at lower $p_T$ regions, the K-factor can reach about 3. In Fig.\ref{fig4}(b) and (d), we present the hardest jet(j1) $p_T$ and $y$ differential distributions. We find that the K-factors are about 2 at all regions. The hardest jet(j1) transverse momentum distribution exhibit a plateau extending up to half the vector mediator mass, which is due to that it could alternatively
originate directly from the vector mediator decay, the similar behavior also occurs in Ref.\cite{Fuks:2016ftf}. For the rapidity differential distributions for the hardest b-jet(b1) in Fig.\ref{fig4}(c), the NLO QCD correction is very large at high rapidity. In Fig.\ref{fig4}(e), the missing $E_T$ distribution is displayed, we can see that the NLO QCD correction effect is not obvious, especially at large $E_T^{miss}$. Due to the uncertainty of numerical integration, the K-factor is not stable at large $E_T^{miss}$ region. For the differential distributions of the distance in the $(\eta, \phi)$ plane $R$ between the hardest jet(j1) and hardest b-jet(b1) in Fig.\ref{fig4}(f), we find that there is a bump from 0.7 to 1.5, and the NLO correction enhance the LO results significantly at this region.

\subsection{Discovery potential for the $X_D \overline{X}_D + b $ production at the 13TeV LHC}
\begin{figure}
\centering
\includegraphics[bb = 25 225 500 550,scale = 0.5]{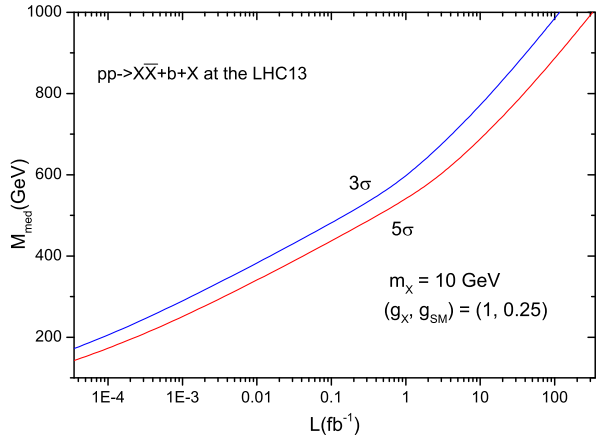}
\caption{\label{fig5} 5$\sigma$ discovery and 3$\sigma$ exclusion limits for the process \ppkkb~ at the 13 TeV LHC with the coupling parameters $g_{X}=1$ and $g_{\rm SM}=0.25$. If a discovery is made, then the regions below the red line is favored. If no signal is found, then the region above the blue line is excluded. }
\end{figure}

In this section, we study the discovery potential for the signal of $X_D \overline{X}_D + b $ production at the 13TeV LHC. In general, the cross section for the process \ppkkb~ is a function of the relevant couplings, the DM mass and the mediator mass. We know that there are more advantages in looking for relative light dark matter particles on the colliders than on the direct and indirect experiments.
From Fig.\ref{fig2}, we can see that the total cross sections have barely changed in the range of light mediator mass, and they are almost same for the vector and axial-vector mediators. Thus, we choose vector mediator mass $m_{X}=10~GeV$ as the characteristic parameter.
The significance of signal over background $S$ is defined as
\begin{eqnarray}
S = \frac{N_S}{\sqrt{N_B}} = \frac{\sigma_S \sqrt{{\cal L}}}{\sqrt{\sigma_B}},
\end{eqnarray}
where $N_{S,B}$ and $\sigma_{S,B}$ are the event numbers and cross
sections for signal and background, and ${\cal L}$ denotes the
integrated luminosity. The SM background mainly comes from the  processes
$p p \to Z + b $ and $p p \to Z + \bar{b} $ $~(Z \to \nu \bar{\nu},  \nu = \nu_e, \nu_{\mu}, \nu_{\tau} )$,
where the neutrino is also the missing energy. Using the same selection cuts as the signal, we obtain the LO cross section at 13TeV LHC as:~
$\sigma = 125.2\pm0.1232$ (pb). If considering the NLO QCD correction, the total cross section is equal to the LO cross-section multiplying a K-factor about 1.25 \cite{Campbell:2003dd}.
In Fig.\ref{fig5}, we present the 5$\sigma$ discovery and 3$\sigma$ exclusion limits for the $X_D \overline{X}_D + b $  production at the 13 TeV LHC. If no signal events will have been detected after accumulating an
integrated luminosity of 116 $fb^{-1}$, then the region with mediator mass $M_{Y} < 1000 GeV$
can be excluded at the 3$\sigma$  level. On the other hand, if the mediator mass satisfies $M_{Y} < 1000 GeV$, then the signal
$X_D \overline{X}_D + b $ is going to be discovered at the 5$\sigma$ level before accumulating an integrated luminosity of
$323 fb^{-1}$. This result shows that there is a more strong potential ability
to distinguish dark matter particles than the mono-$Z$ productions \cite{Neubert:2015fka}.

\vskip 5mm
\section{Summary}
\par
The LHC provides an ideal facility to search for DM particles.
Accurate and precise predictions for DM associated production
rates and distributions are necessary to obtain robust
constraints on different DM models. In this paper, we calculated a pair of fermionic dark
matter particles associated production with a b-jet at the LHC,
including next-to-leading order(NLO)
QCD corrections and parton-shower effects. We have considered
a simplified model where DM is a Dirac fermion
and couples to the SM via either a vector or axial-vector
mediator. For the $X_D \overline{X}_D + b $ production in the vector and axial-vector mediator models, our results
show that higher-order corrections have a significant
effect both on the overall production rate as well as on the
shape of differential distributions. The NLO QCD
corrections to the LO production rates can be very large, the
$K$ factors can reach to 3 in reasonable parameter space. This shows that the NLO
corrections have a noticeable impact on the DM $+$ b-jet signal and must be considered.
We also considered the discovery potential for this process, and found that this process has potential to be detected at the LHC.

\section{Acknowledgments}
This work was supported by the National Natural Science Foundation of China (No.11205003, No.11305001, No.11575002).

%-------------------------------------------------------------------------------------------------------

\end{document}